\documentclass[12pt]{article}
\begin{document}
\begin{center}

\centerline{\bf\large Remembering Nikolai Nikolaevich Bogoliubov}
\bigskip
                   D. V. Shirkov
\end{center}

\tableofcontents

\section{Personal impressions}

\subsection{The Late Forties}

  My first impression dates back to the spring of 1947 when N.N. read
 a special course of lectures on dynamic equations of statistical
 physics. Recall that Bogoliubov became a professor of Moscow State
 University in 1943 after returning from Ufa where the Ukrainian
 Academy was evacuated during the war. In this period he shared his
 time between the Institute of Mathematics in Kiev and the Department
 of Physics (Fizfak) in Moscow University. At the end of 1947 N.N. was
 awarded the Stalin Prize for two treatises on theoretical physics, one
 of which was monograph {\sf Dynamic Equations of Statistical Physics}.

  Rather short, in an elegant grey suit and a bow tie, portly and in
 his prime, lively and buoyant, he enthusiastically lectured following
 in general the above-mentioned book that had been published shortly
 before that. It was obvious that he enjoyed both the subject of the
 lecture and the contact with students. It was somewhat unusual, beyond
 the slightly aloof manner of lecturing maintained at the Fizfak at
 that time, which was impressive in itself and evoked a sort of liking.

  At first, the subject did not seem interesting to me (shortly before
 that, Smith's report on the atomic bomb test had been published in
 Russian, and my imagination was occupied with "deeper-concealed"
 mysteries of the universe); however, the personal charm of the young
 (short of forty years old) and already known professor, corresponding
 member of the Academy of Sciences, and his clear and precise style
 made their effect and I listened the entire course.

  At the end of the next year my student-fellow Valentin Nikolaevich
 Klimov (with whom we worked side by side under N.N.'s supervision for
 about five years and who later tragically died in the snow avalanche
 at the Caucasus) told me that N.N. had got a small theoretical
 department at the Academy Institute of Chemical Physics (IChP) and
 graduate students engaged on their diploma work are needed.
 {\it\small[
 By that time, following the advice of my older friend Yura Shirokov,
 I had been in this position under Dmitrii Ivanovich Blokhintsev for
 about half a year, but was disappointed by my status. At that time
 D.I. was the head of the then secret project for construction of the
 atomic power plant in Obninsk and only occasionally came to Moscow.
 To meet him required time to be spent for telephoning and quite plenty
 of persistence. For all that time I had not been given any problem to
 deal with except a simple test exercise. Thus,]} \\
 I agreed to write my graduation thesis under Bogoliubov's supervision
 without much hesitation.

  It turned out that in addition to the theoretical department that had
 long existed at the IChP and was headed by Prof. Aleksandr
 Solomonovich Kompaneets, another one related to the Atomic Project was
 recently established there. After I came to the department, it
 comprised N.N., Boris Valentinovich Medvedev, and two laboratory
 assistants, Valya Klimov and me.

  Here, coming to my mind is a scene that took place in the office of
 Academician Semenov, director of the institute, later a Nobel Prize
 laureate. N.N. came to him with Klimov and me to make arrangements for
 our official position at the IChP. We were supposed to have a half-pay
 part-time job of laboratory assistants. However, the invited personnel
 manager said that it would take quite a lot of time to go through the
 formalities of registration of the university students as part-time
 workers in Academy of Sciences: it was necessary to apply to the
 Presidium of the Academy, then to the All-Union Qualifying Committee
 to which Moscow University was subordinate at that time, then to agree
 upon the matter with the university and Fizfak administration, etc;
 finally, in the case of positive resolution, the papers were to go
 back in the same long way. Then, after a minute's common confusion
 there came a question from N.N.: -- ``Well, and if we take them as
 full-time workers?'' \ This option turned out to raise no formality
 problems for the personnel department and no objection from Semenov.
 And by the order of the director two young pikes were immediately
 thrown into the river.

  Our theoretical department occupied one room of moderate size. In
 the middle and opposite to each other there were two desks. Two sofas
 served to seat visitors and occasionally to allow the hosts to have a
 nap. An indispensable accessory was the tea making and drinking outfit.
 There were also chess and a chess clock. For any visitor to find us
 working hard, the entrance to the room was, due to efforts of B.V.,
 supplied with a double door with a small space between them. For the
 ``sake of secrecy'', both doors were always locked, and while one of
 us was unlocking them to a knock from outside, the other was clearing
 away the chess and cups from the table.

  The security regulations also implied that the scientific creative
 work should be over no later than 17:45 because all calculations,
 including preliminary rough ones, must be carried out only in tied
 sealing-waxed notebooks which were handed in at the security office.
 Nevertheless, the most fruitful time was the evening, when we were not
 disturbed by the neighbor scientists or inspection raids of fire or
 security officers, etc. We often sat late up to the last trolleybus.
 N.N. showed completely quiet attitude towards our chess playing
 (though he did not play himself) and liberal behavior. He appreciated
 the working qualities and the results obtained.

  The chief assigned me the task to simplify the kinetic transport
 (i.e., neutron diffusion and slowing-down) equation. This beastly
 awkward integro-differential Boltsman equation for the distribution
 function involves three independent variables even in the spherically
 symmetrical geometry. In the general case it allowed only tedious
 numerical calculation. The known approximations (one-velocity,
 diffusion, age-diffusion) were too rough for dealing with real
 problems under consideration.

  From my present-day point of view, the remarkable fact is that N.N.
 only formulated the problem for the student without making even a hint
 at any lines of attack. The problem in question was interesting
 technically and very important in essence: any serious advance allowed
 a hope for appreciable economy in numerical calculation, which led to
 a gain in time.

  At the time before the advent of computers, solutions to complicated
 equations were carried out by numerical calculations with desk-top
 electromechanical calculators obtained from defeated Germany as
 reparations. These machines were usually operated by girls united
 into a calculation team supervised by professional mathematicians.
 The latter prepared difference schemes suitable for paralleling,
 analyzed solution stability, degree of accuracy, etc. Calculation
 pools like that could by no means be found in many institute;
 calculation of more or less complicated problems was expensive and
 took much time. And the factor of time heavily weighed upon our
 activities. The first Soviet atomic bomb was tested only in August
 of the next year, 1949.

  Within a few months serious advances were made in solving the problem.
 I used simplification of the integral operator kernel, the so-called
 scattering indicatrix, as a basis for the new approximation. The main
 idea came to my mind during the Moscow University Komsomol (Young
 Communists' League) conference. Having specially taken my seat at
 the gallery, as far from other physics department delegates as
 possible, I was deep in thought amidst the murmur of the reporting ... .

 Details omitted, it can be mentioned that in the mid-1950s, when the
 pure theoretical part of my investigations was unclassified, two
 papers on the method of the so-called synthetic kernel in the theory
 of neutron diffusion and slowing-down were published in the journal
 Atomnaya Energiya (Atomic Energy). One of them corresponded to the
 graduation thesis written at the IChP in 1949; the other, involving
 the generalization to a more complicated case of neutron transport
 in media containing hydrogen nuclei, to the candidate of science's
 (Soviet and Russian PhD equivalent) thesis defended in May 1953.

  About ten years after, both papers were fully reproduced in the
 monograph by Davison. The method of approximate scattering kernel
 transformation was the subject of the chapter ``Shirkov's Method''.
 I was informed about it at the tennis court in Dubna by Bruno
 Pontecorvo, who, being a student of Fermi, was keeping up with all
 publications in the subject of his great teacher that came to the
 JINR library. It became clear that the American colleagues failed to
 come up with anything equivalent to it. Availability of powerful
 computers panders to the ``have a computer, needn't have the wit''
 philosophy. This collision of the Russian native wit with the spoiled
 American theorists took place again in the mid-1970s; it involved my
 students and concerned the calculations of three-loop Feynman diagrams
 in gluodynamics.

  Along with this, so to speak, major activity I began attending N.N.'s
 seminar at Steklovka (Steklov Institute of Mathematics, Academy of
 Sciences of the USSR) accommodated in a lavishly glazed building
 slightly protruding into Leninskii Prospekt (Lenin Avenue) right
 opposite to what was the Academy Presidium at that time. The seminar
 was held once a week, and when N.N. was absent, it was conducted by
 Sergei Vladimirovich Tyablikov. Among the things studied at the
 seminar was, for example, Schwinger's known series of papers.

  An extremely helpful tradition at the seminar was the review of
 publications. At the end of each meeting the head of the seminar
 looked through a recent issue of a journal, like Soviet Journal
 of Experimental and Theoretical Physics (JRTPh) or American Physics
 Review, pointed out interesting articles, and gave them out to the
 young colleagues for abstracting. At the next meeting, the main
 talk was preceded by one or two five-minute essays on the previously
 given topics.

  This system yielded two results: first, all participants were
 regularly briefed on the news; second, the audience was not divided
 into those active and those passive. If you started attending the
 seminar, kindly work and show with your essay what you know and how
 critical you can be about somebody else's results. My first essay was
 about the "sensational" statement that there existed classically
 stable electron orbits around the positively charged nucleus, which
 was published in the Physical Review. The mistake was that the
 quadrupole and higher multipole radiations have been neglected. That
 essay with the error analysis enhanced my status among the seminar
 participants.

  It should be mentioned that at that time, in the late 1940s, N.N.
 had just turned from statistical physics to theory of particles.
 (These turns in the subject of studies were typical of Bogoliubov
 when he, having ultimately solved a complicated problem, lost interest
 in that topic for ever). His first papers on the covariant
 formulation of the Schr\"odinger equation were published in 1951.

  The seminar was held at the end of the workday; then N.N. and the
 participants went out, and after walking about half a kilometer
 along the street the company turned to a shop under the signboard
 ``Ararat''. There, one would have not only a bottle of Armenian
 cognac uncorked but also glasses and sliced lemon served. And that
 was the actual finale of the seminar.

  In late 40s, N.N. lived with his family in Kiev and regularly came
 to Moscow staying at the Moskva or Yakor (in Gorky Avenue) hotels.
 His arrivals and departures were like small festivities celebrated
 in restaurants where N.N. invited all his co-workers, including
 students. Generally, and especially in those younger years of his,
 N.N. was a very cheerful and actively amicable man. He liked to
 enjoy life and share this joy with others.

  Two strong impressions of N.N.'s personality at that time (to the
 student's eye): devotion to his science and a high level of culture.
 Scientific work seemed to be the major point of his life and the main
 source of joy. He did not play chess or cards and did not go in for
 sports. For him, a good pastime meant good intellectual work. A
 relevant reminiscence from the 1960s: to my question, -- ``Did you
 have a good rest?'', which I asked N.N. after his coming back from a
 Caucasian sanatorium, there followed -- ``Yes, excellent. Two works
 are finished.''

 Contacts with N.N., giving rise to a liking and an involuntary wish to
 imitate, led to rethinking of life priorities values -- intellectual
 activity did not merely moved to the first place, it took top
 seniority. 

  N.N.'s immense erudition in literature, history and linguistics,
 amazed me, a well-read boy from a professor's family. Those
 impressions aroused continuously, got embedded in serious scientific
 discussions, were enhanced by some satirical chords. His wise, calm,
 and somewhat ironical attitude towards life was based on, so to speak,
 stable invariants formed in his youth. Though N.N. never spoke about
 religion, his moral rules, conveyed to the students in some hidden
 ways, complied with the Christian commandments.

 When cited by him, eternal motives and characters produced by classic
 authors got implicit both in quite simple psychological situations
 and in unexpected turns of world policies\footnote{Based on the
 experience in contacting with a lot of eminent scientists gained for
 more than 50 years of my academic life, I can now add that this
 impression of N.N.'s intellectual and moral exclusiveness has only
 enhanced with time.}.

  The described period ended in the spring of 1950, when our group
 was transferred from Moscow to where, according to a graphic phrase,
 ``Khariton\footnote{Prominent physicist, the chief designer of Soviet
 nuclear weapon.} drove the calves''\footnote{ The full verse by A. S.
 Kompaneets: ``I will be quick as a flea // and slippery as a triton //
 To keep away from places // where drives his calves Khariton''.}.

 \subsection{At the Installation} One fine day in March, I was called
 to the IChP security department and informed that I was to be
 transferred from the institute to ``Lab-2''. This lapidary name was
 given to what is known now as ``Kurchatnik'' (Kurchatov Institute).
 However, at that time Lab-2 disguised the PGU -- First Chief
 Directorate at the Council of Ministers of the USSR, which was in
 charge of the atomic matters and was later transformed into Sredmash
 (Ministry of Medium Machine Building of the USSR). In the personnel
 department of this institution I was flabbergasted at the news that
 I was to leave for the new place of work within a week. I was not
 informed about either the name of the place or how far it is from
 Moscow.

  A departure to the Installation (the colloquial name for our town at
 that time, ``obiekt'' in Russian) proceeded as follows. The personnel
 department of the Large House sent the novice to one of the central
 Moscow squares where he was ``to enter the unlighted arch of building
 No. XX, turn to a shabby door without a sign, go several meters along
 a narrow corridor in utter darkness, grope for a door on the left and
 open it''. On fulfilling all these instructions, you found yourself
 in a lamp-lit room standing in front of a man who sat at the table
 and immediately addressed you by name as if you were acquaintances.

  He handed me a travel document and directed that the next day I
 ``arrive at the Vnukovo airport with the belongings, sit on an
 indicated bench in the waiting hall at an indicated time''. The
 answer to all my questions about any details or explanations was,
 ``I can't tell you that, but don't worry''. The next day an unknown
 man came up to me in Vnukovo at the assigned time, called me by name,
 and told me that boarding was beginning in a few minutes and I should
 follow him. A quarter of an hour later, without any announcement of
 the flight, a group of people led by a guide crossed the airfield
 without passing the ticket control and embarked a twin-engine,
 apparently 12-seat cargo version of the Douglas aircraft with
 aluminum seats along the sides, which immediately taxied to the
 runway. Owing to fair weather at the time of landing, I could
 approximately determine the place of destination. On leaving the
 plane I caught the sight of control officers in the KGB uniform and
 \ ... \ Valya Klimov, who came to airfield to meet me.

  The term \ ``Zvonkovoe'' \ was proposed by N.N., who took it from an
 operetta popular in the pre-war time. A byword from it, \ ``Come to
 see us in Zvonkovoe'', was particularly in place for our top-secret
 small town situated in a vast forest and surrounded with barbed wire
 fences.  N.N.'s group in \ ``Zvonkovoe' \ originally included Klimov
 and me. Quite soon Dmitrii Nikolaevich Zubarev joined it, and in 1951
 there came Yurii Aleksandrovich Tserkovnikov and Vasilii Sergeevich
 Vladimirov.

 In terms of work, our group was closely linked with the group of Igor
 Evgenievich Tamm from FIAN (Lebedev Physics Institute). Both teams
 were simultaneously transferred from Moscow to the Installation in
 the spring of 1950 for intensification of the work on the
 hydrogen bomb constructing.

 In terms of everyday life, our two teams were a single whole.
 Beginning with the autumn of 1950 several theorists without families
 (including those whose families stayed in Moscow) were lodged in a
 two-storied two-flat cottage built to the standard of the atomic
 agency as of the late 1940s. Similar cottages have still remained not
 only in Sarov but also in Dubna, the Chernaya Rechka (Black River)
 area. Upstairs, in the two-room flats of each half of the cottage,
 there lived the Corresponding Members of the Academy Igor Evgenich
 (I.E.) and N.N. Downstairs, under I.E., one room was occupied by
 Andrei Dmitrievich Sakharov, who was replaced, after his wife Klava
 with daughters came, by Tserkovnikov, and in the second room lived
 a young theorist Yura Romanov; in the other half-cottage, under
 N.N., -- Valya Klimov and me.

  The residents of the theoretical cottage together with a few young
 theorists from I.E.'s and N.N.'s groups who resided in the nearby
 hotel set up an informal household association \ ``United Theorists
 Organization''. The UTO members have common board: non-resident cooks
 -- elderly and cheery Aunt Sonya and younger Valya -- made and served
 breakfast and lunch. Young people stored up food. Wholesale purchases
 were made using the ``Pobeda'' car attached to the theoretical
 department, which usually brought us to Institute a couple of
 kilometers away by nine o'clock in the morning and back home for
 lunch in the afternoon.

  On some days off (at that time a week was a six--day period and each
 day that was a multiple of 6 was a rest day) we went for wholesale
 \ ``foraging'' \ to a large village of Diveevo (to the church of
 which relics of Serafim Sarovski were moved in the post-Soviet time)
 situated outside the Zone. To leave the Zone, one had to get a pass
 for the entry control point. The Zone was an area of several hundred
 square kilometers surrounded, in full compliance with the penitentiary
 camp norms and standards, by barbed-wire fences, an exclusion zone,
 watchtowers, searchlights, etc.

 Holiday markets were abundant in goods, the customers from the Zone
 were lavish, and it was rumored among the local people that there, \
 ``behind the barbed wire'', an experiment on building communism was
 going on.

  The chiefs, i.e., I.E. and N.N., liked to sleep longer in the morning
 and usually came to office by eleven o'clock or so. N.N., who spent
 about half of his time at the Installation, made regular reports to
 review the news in \ ``unclassified'' \ science, mainly in quantum
 field theory. It was remarkable that immediately after lunch N.N. --
 40 years old at that time (!) -- had a customary rest, \ ``to wash the
 brain'', as called it. After 4 p.m. music was heard from the radio
 above. It meant that the chief got up, had tea, sat down to work. At
 that time he was already allowed to be disturbed.

  The actual working hours of the theorists were regulated by the
 security department. The work on the main topic, including rough
 calculations and drafts, could only be done in special personalized
 tied and sealing-waxed notebooks slightly larger than the A4 size
 with each page numbered. Each of us had a special briefcase to carry
 the notebooks and a personal seal with a number. The briefcase was
 either with his owner or, sealed by him, in the security department.
 It could be taken from or handed to the department only within the
 working hours. Unlike the case in the IChP, it was prohibited to stay
 in the working room in the off-hours. Only sometimes, before a regular
 test at the far testing ground, an all-out effort order would be
 issued for some departments. Therefore, after 6 p.m. and on days-off
 it was possible to think over and discuss working topics only during
 the walk in the forest after making sure that you could be heard by
 nobody except birds.

 Under these conditions, it was quite natural to
 be occupied with unclassified science in the evenings, especially
 under continuous influence of such figures as Tamm and Bogoliubov.
 It was just at those years that I began seriously studying quantum
 field theory in my off-hours.

  A couple of illustrations to the psychological portrait of N.N. I
 remember the day when the Korean War began (25 June 1950). It was
 reported in the morning news that the troops of South Korea, the
 United States' satellite, had suddenly crossed the 37th parallel,
 which was the state boundary at that time, treacherously invaded
 peaceful democratic North Korea, broke the frontier defense line, and
 moved several tens of kilometers into the area. However, the valiant
 North-Korean army succeeded in regrouping, overthrew the aggressor on
 the same day, and carried the war into the enemy's territory.

  On that day N.N. came by air from Moscow. Valya and me met him.
 In the car with N.N. on the way from airfield, I am excitedly telling
 him all this official propaganda rubbish. As if he does not hear
 anything, N.N. starts telling us what is new in Moscow. We come home
 and begin helping him with his luggage. N.N. puts the kettle on the
 stove and turns on the radio. Here comes another summary of victories
 gained by the North Korean army. And the chief's face suddenly gets
 distorted as if by pain… It becomes clear that in the car he took my
 words for an attempted practical joke and probably wondered inwardly
 at our intellectual awkwardness. And it turned out that it was not
 the rubbish but the reality the responsibility for which rests with
 much higher-ranking persons who govern our way of life.

  N.N. did not like abundant words. Thus, the paraphrase \ ``First
 think, and then speak'' \ of the remark which Tamm put into Dirac's
 mouth, \footnote{In full, the remark made by Dirac to Niels Bohr, of
 which
 I.E. was a witness, is as follows: "In my childhood, my mother taught
 me first to think and then to write".} is fully applicable to him.
 This is well illustrated by the above episode. Another man of the
 similar kind was the scientific supervisor of the Installation and
 the chief designer of nuclear weapons Yulii Borisovich Khariton. Any
 time when N.N. needed help or advice of Yu.B., he went to see the
 latter and gave him the gist of the matter. As a rule, Yu.B. never
 gave an immediate answer. After a short pause he changed the topic.
 At another meeting in a few days he could return to the problem and
 propose a solution. And could not, either. Like a Japanese, who
 avoids saying \ ``No''. To characterize the procedure of inputting
 information into Khariton's brain and gradually arriving at a fully
 shaped solution, N.N. used the verb {\it to kharitonize}.

 Finally, one more episode, now from the 1970s\footnote{Cited according
 to the witness, a student of Bogoliubov.}. The scene is laid in the
 office of the JINR director. The secretary informs N.N. that an
 Academician X, director of one of the JINR Laboratories, has suddenly
 come to see him. The Academician enters and effusively explains that
 a major discovery recently made at his Laboratory did not win the
 recognition of Western colleagues. He proposes to discuss the
 situation at the nearest meeting of the Scientific Council of JINR
 and ask the Council to take a decision as to the reality of the
 discovery. N.N. listens to his emotional interlocutor with perfect
 calm and ... offers a cup of tea. While having tea, he speaks about
 some scientific news. Tea drinking is finished, and X returns to his
 problem. Then N.N. says, \ {\small\it``I have just tried to imagine
 that the director of the Mathematical Institute Academician Vinogradov
 makes a proposal to the members of the Scientific Council of Steklovka
  Kolmogorov, Pontryagin, Aleksandrov ...  to consider this or that
 theorem proved ... ''} \
  Without listening to the whole, X rushes away from the office.

 \subsection{Bogoliubov and Lavrentiev}

 I think of my quite intimate, almost family-like acquaintance with
 two remarkable people, Bogoliubov and Lavrentiev, as being the
 greatest heaven's gift.

  In outward appearance, Nikolai Nikolaevich and Mikhail Alekseevich
 were a rather contrasting pair. Somewhat portly, medium tall N.N. and
 thin, very tall M.A. A handsome face of N.N., with slightly wavy thick
 hair even in his declining years, and a rather long face of M.A., with
 scanty hair. ``Outwardly clumsy, sometimes even awkward'' \ (as Boris
 Evgenievich Paton put it) Lavrentiev and dandified and elegant,
 artistically looking, often in a bow tie, Bogoliubov. Similar in
 their appearance were large foreheads and the expression of the
 serious eyes.

 They became friends in Kiev as far back as the 1930s when the ten-year
 difference in age was yet significant. Mikhail Alekseevich had known
 Bogoliubov's teacher N.M. Krylov well and used to call tenderly his
 friend Kolyasha behind his back. In science-like conversations N.N.
 often gave examples and episodes which involved Mikhail Alekseevich,
 whom he loved and esteemed. As a result, without seeing him, I had
 already a myth of M.A. formed in my mind.

  And there came the time we met. It was in Sarov in May 1953. I just
 returned from Moscow where I got the PhD degree in the Scientific
 Council of Lab-2 presided by Igor Vasilievich Kurchatov himself.
 Accordingly, a small banquet was given in our theoretical cottage to
 celebrate the successful defense of the thesis. We had already drunk
 a couple of toasts when the chief, who was a bit late, said while
 sitting down to table, \ -- ``Lavrentiev has come to our place''. To
 my reply, -- ``It would be nice to invite him'', \ there followed,
 --``No problem, here he goes away from our house.'' I rushed out,
 came up with Mikhail Alekseevich, introduced myself, and immediately
 invited him to our place. He agreed without hesitation and we came
 back together.

  As was already said, I had worked about three years in Sarov by that
 time. That period was associated with the development of Sakharov and
 Tamm's \ ``puff'', for the participation in which I was given my
 first state award.

  The innocent acquaintance at the celebration table entailed serious
 consequences. In the autumn of 1953, after finishing the work on the
 puff, N.N. (as well as Tamm) returned to Moscow having \ ``handed''
 me to Lavrentiev, in whose team I worked on the nuclear filling of
 the atomic artillery shell for the next three years.

 Scientifically and technologically, the problem was to turn a physical
 spherically symmetrical structure (the first American bombs thrown on
 Hiroshima and Nagasaki, and the first Soviet ones as well), a sphere
 short of one meter in diameter with some 10 kg of uranium-235 or
 plutonium, into a sort of melon with the cross dimension allowing it
 to go in a cylindrical shell of caliber no larger than 400 mm.

 The breaking of the spherical symmetry make is appreciably more
 difficult to calculate the now non-synchronous ignition of the
 detonators, the hydrodynamics of the convergence of the implosing
 shockwave to the center of the article, and the process of
 development of the nuclear chain reaction. 

  The joint work with Lavrentiev, which ended in a successful test at
 the Semipalatinsk testing ground in March 1956 and was crowned with
 the Lenin Prize, resulted in close contacts with M.A. within the
 second half of the 1950s. On returning to Moscow, Mikhail Alekseevich
  devoted himself to a new grandiose patriotic activity
 of establishing the Siberian Branch of the Academy of Sciences. It
 was already at Sarov that he began looking for those who
 would help him with developing Siberia. In the late 1950s, working
 at Steklovka and in Dubna, I several times traveled on business to
 Novosibirsk and the site of future Akademgorodok. Nikolai Nikolaevich
 rendered his support to the Siberian project by participating in the
 Commission of the Presidium of the Academy of Sciences for the
 establishment of the Siberian Branch.

  In 1958 M.A. acquainted me with one of his major associate Sergei
 Lvovich Sobolev, who started the Institute of Mathematics in
 Novosibirsk, and offered me to become the head of the Theoretical
 Physics Department at that institute. I began selecting the staff.
 At the first election of Academy members for the Siberian Branch in
 1958 I was included in the Corresponding Member ballot list, but
 success came only at the second election two years later when I was
 jointly nominated by Academicians Bogoliubov, Lavrentiev, and Sobolev.

  In the autumn of 1960 I moved to Akademgorodok near Novosibirsk.
 One of the first vivid impressions was celebration of Lavrentiev's
 60th birthday in unexpectedly cold November with severe frosts.
 Nikolai Nikolaevich was among the guests.

  Lavrentiev could influence people and get them into his activities.
 Nikolai Nikolaevich used to say that M.A. is \ ``a master of playing
 chess, human chess''. To implement his plans, e.g., in establishment
 of the Siberian Branch, Mikhail Alekseevich looked for those who
 shared his ideas and was willing to help among the professionals in
 various fields of science, art of organization, journalism, civil
 engineering ... . Among them, he especially valued those who were
 similar to him in the main feature-service to the cause. And he
 treated them as friends.

 \section{Joint work}

 \subsection{Quantum Field Theory}. Nikolai Nikolaevich began to
 study hard the problems of quantum field theory (QFT) at the end of
 the forties undoubtedly under the influence of the well-known
 papers by the founders of the modern covariant field theory. Anyhow,
 his first quantum-field publications appeared in 1950-51, three of
 them being devoted to equations in variational derivatives of the
 Tomonaga--Schwinger type. The latter were based on the axiomatic
 definition of the scattering matrix as a functional of the Bogoliubov
 function of the interaction region $g(x)\,$ standing for the
 Schwinger surface function $\sigma(x)$. In the first half of the
 fifties, N.N. was actively involved in the rapidly developing field,
 the renormalized quantum field theory. He entered it on
 the side of mathematics, nonlinear mechanics, and statistical physics,
 already having results of a world level. He went into it longer and
 grew into it deeper than other scientists migrating to QFT from
 mathematics (Gelfand) and other more classical fields of theoretical
 physics. To some extent N.N. trajectory was similar to that one of
 the English--American theoretician Freeman Dyson. It is known that
 Bogoliubov created his renormalization method on the basis of the
 Sobolev-—Schwarz theory of generalized functions. Recall that the
 Bogoliubov renormalization method was appearing approximately at the
 time of writing our book -- in the mid-1950s. Nikolai Nikolaevich and
 his fellow workers (Ostap Stepanovich Parasyuk and later V.S.
 Vladimirov) were led to considerably elaborate the works by Sobolev
 and Schwarz as applied to QFT needs. In particular, to introduce a
 class of functions that made possible the Fourier transformation and
 determination of the procedure of multiplication of singular functions.
 Within his approach one does not need to introduce ``bare'' fields
 and particles and can do without a physically unsatisfactory picture
 of infinite renormalizations.

   Nikolai Nikolaevich used to give talks from time to time at the
 Sarov theoretical department with the review of large parts of novel
 QFT such as ``renormalization'', ``functional integral'', or ``surface
 divergences''. Listeners to these reviews were impressed by that N.N.
 ``saw'' those so different fragments from a single point of view and
 perceived them as part of one picture. Recall that that was the time
 when textbooks on particle theory were only pre--war editions of
 ``Quantum Theory of Radiation'' by Heitler and the book by Wentzel
 issued at the beginning of the forties. ``Quantum electrodynamics''
 by Akhiezer and Berestetsky (1953) as well as the first volume of
 ``Mesons and fields'' by Bethe, Hoffman and Schweber (1955) were
 awaiting for their appearance.

 One day in the autumn of 1953, being impressed by one of his lectures,
 I asked, -- ``Nikolai Nikolaevich, Why do not you write a book, a
 textbook, on QFT ?'' The reply was, -- ``Not a bad idea. Probably, we
 could realize it together ?''.  First, I did not take this suggestion
 seriously. The list of N.N.'s papers shows that by his 25th
 anniversary N.N. was the author and coauthor of several monographs,
 whereas for me it was something new. However, not only bad habits
 are contagious and ten years later my disciples Ginzburg and
 Serebryakov, who were under 30, became coauthors of the book on
 dispersion relations. Later this situation was repeated with Belokurov.
 To justify my reaction, I should like to note that only in May of
 that unforgettable year one of the coauthors of a future book had
 defended his Candidate Sc. Dissertation (PhD) on the theory of neutron
 transfer and had none of the papers on quantum field theory, whereas
 the other became Full Member of the Academy of Sciences in October.

  However, in a week the conversation was resumed and we began to
 discuss the details of the project. The time frame of those events
 is reliably determined, as  by the time we sent a proposal to
 Gostekhizdat at the beginning of 1954 the book by Akhiezer and
 Berestetsky had just been published. Besides, the first
 version of the consistent presentation of the Bogoliubov axiomatic
 $S$ matrix was submitted for publication in Uspekhi Fizicheskikh
 Nauk at the end of 1954.

  The first draft of the book, except for the introductory part
 expounding the Lagrangian formalism of relativistic fields and the
 Schwinger quantization scheme, included an original axiomatic
 construction of the scattering matrix mainly built on the Bogoliubov
 causality condition, his renormalization method based on the
 distribution theory, as well as the method of functional integral and
 the generalized Tomonaga—Schwinger equation.

  The work on the book followed the scheme \ ``gasoline is ours – ideas
 are yours''. Most part of the work was done at N.N.'s home in the main
 building of Moscow State University (MSU) on the Lenin hills where we
 talked for an hour -- two making the outline of yet another section.
 After that I wrote the first version of the text that was discussed
 at our next meeting and very often was considerably revised. The
 thoroughly rewritten manuscript, if finally approved by the chief,
 was put in the upper left corner of the big wardrobe from to be taken
 and typed by his wife, Evgeniya Aleksandrovna. Slightly embossed paper
 of several colors was used for typing. This type of paper, produced
 by the Riga factory, was specially bought for our work. Nikolai
 Nikolaevich liked it very much. Different paragraphs of the typescript
 had different colors: blue, yellow, light-green, violet ... . Three
 copies were typed at a time. I took the typed paragraphs from the
 opposite right corner of the wardrobe to write down the formulae.

 The third copy of colored paragraphs stitched into chapters was
 intended for critical reading by collaborators of N.N.'s department
 in the Steklov Institute. This reading was the first \ ``running in''.
 Two extensive papers submitted to Uspekhi were meant as the second
 one. Therefore, the text of the book issued in 1957 was for the most
 part rather well \ ``ironed'', and except for the fresh material of
 the last two chapters on renormalization group and dispersion
 relations it displayed, in a sense, the \ ``third approximation''.
 Looking back from the vantage point of my subsequent experience, I
 must say that this monograph over 30 printer's sheets long, was
 written rather quickly. The decisive reason, in my opinion, was that
 N.N. had a clear--cut plan in his head from the very beginning, and
 kept in his mind the whole written text afterwards.

 \subsection{\small The birth of the Bogoliubov renorm-group}
 In the spring of 1955 a small conference on ``QED and
 elementary particle physics''was held in Moscow. It took
 place in the Lebedev Physical Institute (FIAN) in the first
 decade of April. Among the participants, for the first time
 in the post-war period, there were several foreigners,
 particularly the well-known theorists Ning Hu from China and
 Gunnar K\"allen from Sweden. My short presentation concerned
 consequences of the Dyson finite transformations for
 renormalized Green functions and matrix elements in QED.
 The central event of the conference was the talk ``Basic
 problems of QFT'' by Landau in which he dwelled upon
 ultraviolet (UV) behavior in the local quantum field theory.
 Not long before, the problem of behavior at small distances
 in QED was considerably advanced in a series of papers by
 Landau, Abrikosov, and Khalatnikov. They succeeded in
 constructing a closed approximation to the Schwinger—Dyson
 equations which turned out to be compatible with both
 renormalization and gauge covariance. Their approximation
 admitted explicit solution
 in the massless limit that turned out to be equivalent, in
 modern terms, to summation of leading UV logarithms. The most
 remarkable fact was that the solution appeared essentially
 self-contradictory from a physical point of view, as it
 contained an unphysical (``ghost'') pole in the renormalised
 amplitude of a photon propagator – the problem of ``zero
 charge''. Landau's verdict was pessimistic: forget about
 the local quantum field theory and Lagrangian. Just this
 thesis was advocated by Isaak Yakovlevich Pomeranchuk,
 Dau's coauthor in ``zero-charge'', in a conversation with
 me. In the name of this thesis he even closed his seminar
 on quantum field theory in the Institute of Theoretical
 and Experimental Physics (ITEPh) and recommended younger
 colleagues to change their field of interests. In those
 days our meetings with N.N. were regular and intensive, as
 we were busy with the preparation of a rather advanced text
 of our book. N.N. was very curious of the results of the
 Landau group and posed a task for me to estimate their
 reliability by constructing, for example, the second
 approximation (including, in modern terms, the
 next-to-leading UV algorithms) to the Landau-et-al
 solution of Schwinger—Dyson equations for verification
 of the stability of UV asymptotics and the very existence
 of a ``ghost''  pole.

 At that time I often met Abrikosov, an acquaintance of mine
 since we were students. Soon after the FIAN conference Alexej
 let me know of the just published paper by Gell--Mann and Low.
 The paper dealt with the same problem but, as he said, was
 rather complicated for understanding and difficult to be
 combined with the results obtained by their group. I looked
 through the paper and shortly informed my teacher of its
 method and results that included rather complicated
 functional equations and some general statements of scaling
 properties of the distribution of electron effective charge
 at small distances from its center.

 The scene that followed my report was quite impressive. N.N.
 immediately claimed that the Gell--Mann and Low approach is
 correct and very important, it represented the realization of
 \ {\it the group of normalization} \ discovered a couple of
 years before by  Stueckelberg and Petermann (published in
 French!) while discussing the structure of finite arbitrariness
 in matrix elements that arose after removal of divergences.
 That group was an example of continuous transformation groups
 studied by Sophus Lie. It followed that the group functional
 equations, similar to those derived by Gell-Mann and Low,
 should hold in the general case, not only in the ultraviolet
 limit. Then N.N. added that the most potent tool in the Lie
 group theory was differential equations corresponding to
 infinitesimal group transformations. Luckily, I was acquainted
 with the fundamentals of the group theory.

  Within the next few days I managed to reformulate the Dyson
  finite transformations for electron finite mass case and
  derive the sought functional equations for scalar propagator
  amplitudes of QED corresponding to group transformations as
  well as the relevant differential equations, i.e., Lie
  renormalization group equations. All the derived equations
  contained a specific object -- the product of the electron
  charge squared by the transverse amplitude of the dressed
  photon propagator. We called this product the \ ``invariant
 charge''. From a physical point of view it represents an
 analog of the so-called function of electron effective charge
 first considered by Dirac in 1933 and describing the charge
 screening effect due to quantum vacuum polarization. We also
 introduced the term \ ``renormalization group'' \ in the first of
 our publications in Doklady Akademii Nauk in 1955\cite{6} (and
 Nuovo Cimento in 1956)\cite{7}. In the second simultaneous
 publication\cite{8} (after two line calculations) ultraviolet and
 infrared asymptotics of QED at the one-loop level were
 reproduced which were in agreement with the above-mentioned
 results of the Landau group. Also the novel two-loop solution
 for invariant charge was obtained which made it possible to
 discuss if the problem of \ ``zero charge'' is real.

   \section{Bogoliubov and Landau}

  The relationship between two great scientists is undoubtedly a very
 delicate subject. A lot of different things had long been piled up
 around these relations, both personal and at the level of their
 Schools. Since Lev Davidovich was my first teacher in modern physics,
 I think it appropriate to share my impressions and my insight in how
 these relations developed.

  I will start with the description of the personality of Dau whom I
 got to know in 1946 as a second-year student. After a brief phone
 talk with the famous scientist I was invited to his place for an
 entrance mathematical interview on his well-known theoretical minimum.
 A penetrating and cheerful glance, an aquiline profile and a curly
 forelock, promptitude of speech, quick reaction and agility, as he
 flew upstairs to his study on the second floor, leaving me alone to
 think over the next question, made a deep impression on me. I
 proceeded to the study of \ ``Mechanics'' \ (Landau and L. Pyatigorsky
 were the authors of the first pre-war edition) and started to attend
 theoretical seminars in \ ``Kapichnik''. Dau was very artistic by
 nature, liked and knew how to produce an effective impression.
  Leading the seminar, at which he was head and shoulders above all
 other participants, he did not miss the opportunity to amuse the
 audience by bright {\it mise en scene} immediately going deeply into
 complicated original constructions of the author (of which there
 were sometimes rather well-known scientists) and very often by a
 few remarks making mincemeat of the speaker. I was a witness of a
 scene like that suffered by Gelfand. However, only the closest
 colleagues of Dau knew that to obtain admittance to the seminar's
 platform a pretender had to pass the \ ``purgatory'', i.e., to
 present his work to Dau himself.

  Unlike many eminent theorists of that time, Dau fully realized the
 importance of mathematics for theoretical physics and often used it
 masterly. It is indicative that his theoretical minimum included two
 maths-related subjects as the first thing, including the examination
 on the qualitative theory of differential equations with special
 emphasis on the analysis of singularities. Dau's maxim ``physics
 begins where a singularity occurs'' is related to this.

 \subsection{Three episodes} The first episode happened in October,
 1946 when N.N. reported on his work\cite{9} on the theory of helium
 superfluidity at a general meeting of the Physics--Mathematics
 Division of the Academy. By that time Dau had been a classic of
 superfluidity for five years, the author of the well-known
 phenomenological theory qualitatively explaining the phenomenon due
 to the presence of a collective linear branch of sound vibrations in
 the spectrum, as well as employing  the notion of peculiar elementary
 excitations (torsional vibrations), rotons, introduced by him.
 Participants of the meeting recall that Landau harshly polemized with
 the speaker whose reasonings were based on the physical hypothesis of
 a crucial role of condensate, i.e., an essentially collective effect.
 However, Lev Davidovich quickly digested and evaluated what he had
 heard, as two or three weeks later he submitted for publication a
 short article \cite{10} in which he suggested a curve with flexure for
 a spectrum of excitations. Of an object independent of the
 spectrum of sound excitations the roton spectrum turned into part
 of the whole curve.

 Landau's phenomenological curve results from N.N.'s formula under
 the assumption of the form of interaction of helium II atoms.
 Landau's article finishes with the phrase being a paraphrase of
 Bogoliubov's conclusion made in his report and related publication
 \cite{9}. However, any reference to Bogoliubov's report or paper
 was missing in Landau's short note. True enough, later in a more
 detailed paper \cite{11} (see also \cite{12}) he manifestly
 pointed out Bogoliubov's priority, ``It is worthwhile to note
 that N.N. Bogoliubov, making an ingenious use of second
 quantization, has recently succeeded in determining, in the
 general form, the energy spectrum of Bose -- Einstein gas with
 weak interaction between particles.''

  The second ``round'' took place in 1955 in connection with the
 issue of ``zero charge'' in quantum electrodynamics. Without
 going into details I should like to note that the analysis of
 this problem made by N.N. with the help of his just developed
 renormalization group approach \cite{13} suggested that the
 conclusion of Landau and Pomeranchuk about the internal
 inconsistency of the local quantum field theory had no status
 of a rigorous result independent of perturbation theory. In a
 certain sense, a psychological scheme of the 1946 conflict
 occurred again: a rigorous mathematical reasoning at a deeper
 level led to a more general and precise picture than previous
 semi-intuitive scheme. As is well known, 10-15 years later the
 local Lagrangian perturbation theory recovered its status of
 the basic method of investigations in particle theory. However,
 the rigidity of the prominent physicist's conclusion \cite{14}
 had a serious effect. It hindered the development of QFT and
 entailed some dead-end constructions like the ``bootstrap''
 theory.

  To the most serious test Dau's ambition was put in 1957 when
 N.N. suddenly intruded into the theory of superconductivity.
 The phenomenon of superconductivity, discovered in 1911, had been
 a painful challenge to leading theorists since the late 1920s.
 It was clear that superconductivity was a macroscopic
 manifestation of the laws of quantum mechanics. It was
 intensively studied by experimentalists, though the key point
 to theoretical understanding was difficult to comprehend.
 Dau had been working in this field since the mid-30s and
 together with V.L. Ginzburg constructed in 1950 a
 phenomenological theory of superconductivity on the basis of
 the two-component order parameter.

 An initiating pulse for Nikolai Nikolaevich to start working out the
 theory of superconductivity was a short note by Leon Cooper\cite{L-C}.
 N.N. immediately saw an analogy with the phenomenon of pair
 correlation of boson type that he had discovered in creating the
 theory of superfluidity. Using the Fr\"ohlich Hamiltonian for
 interaction of electrons with phonons (excitations of ion lattice)
 as a basis and modifying his $(u, v)$ - transformation from the
 theory of superfluidity for fermions Bogoliubov applied \cite{15}
 the condition of compensation of possible singularities in the
 vicinity of the Fermi sphere surface and derived an expression
 for the energy gap of the type of the Cooper formula with
 nonanalytic dependence on the Fr\"ohlich coupling constant
 (see below Section \ref{322}).

 At the time, when N.N. finished his investigation and began to talk
 on it at seminars, a preprint by Bardeen, Cooper and Schrieffer (BCS)
 paper was rumoured to appear in the West. However, this preprint did
 not reach Moscow yet. As far as I remember, Dau quickly evaluated
 Bogoliubov's work. It was even agreed upon organizing a joint
 Bogoliubov-Landau seminar on the issue of superconductivity. At the
 first meeting, after N.N.'s talk Dau concluded, \ ``Nikolai
 Nikolaevich, I do not know what there is in the paper by Bardeen and
 others, but I do not think that they have got such a beautiful and
 convincing result like yours.''

 This episode shows that at the time described Dau already considered
 N.N. an outstanding theoretical physicist, having subdued his own
 emotions. The Landau--Bogoliubov seminar existed not long and ceased
 after the appearance of the Physical Review issue with the paper of
 the three authors who proceeded not from the Fr\"ohlich Hamiltonian
 for electrons and phonons, but rather from some approximate model
 Hamiltonian $H_{BCS}\,$ for electrons only postulating effective
 attraction between electrons with opposite momenta and spins in the
 vicinity of the Fermi surface. Dau's sentence turned out to be
 prophetic.

  Nevertheless, mention of Bogoliubov's papers is seldom made by
 representatives of the Landau School in their publications on
 superconductivity, microscopic theory of superconductivity is
 called the BCS theory, and the term ``theory of superfluidity''
 is bound up with the name of Landau only.

 \subsection{Supplementing each other} Spontaneous symmetry breaking
 (SSB) is the subject of the 2008 Nobel Prize in Physics. This topic,
 in a sense, united two outstanding theoretical physicists Bogoliubov
 and Landau by their joint contribution to the explanation of the
 mechanism of phase transformations in large quantum systems followed
 by spontaneous symmetry breaking\footnote{For a more detailed
 discussion of the spontaneous symmetry breaking in quantum physics
 see recent review \cite{60-ufn}.}. The case in point is the systems
 that are described by mathematical expressions having some symmetry,
 whereas a real physical state of the system corresponding to
 particular solution of equations of motion has no this symmetry. A
 situation like this appears when the lowest of the symmetric states
 does not provide the system with absolute energy maximum and is
 unstable. In this case, the particular lowest state is not the only
 one and their set forms a symmetric set. A real reason for symmetry
 breaking and transition of the system to one of the lowest
 nonsymmetric states is arbitrarily small nonsymmetric perturbation.

  For simple illustration turn to classical mechanics. Let us take a
 system consisting of an empty vessel with the concave bottom (a
 champagne bottle) and a small ball. Put this vessel, which is a
 body of rotation, in a vertical position and above it a small
 ball exactly along the axis. A system like this is symmetric with
 respect to rotation along the vertical axis. Let the ball go to
 the bottom. Upon reaching the bottom the ball will not rest on
 the central convexity and will slide to the brim of the vessel.
 Thus, the initial conditions are symmetric, whereas the final
 state is nonsymmetric.

  The initial material of physics, data of observations, has to be
 put in order and interpreted. The way of ordering consists in
 constructing a phenomenological scheme based on a representation,
 a physical picture, of the nature of a phenomenon expressed in a
 mathematical form, a form of a physical law. An important criterion
 for a system and its representations to be successful is not only
 the description of available data, but also a possibility to
 predict results of new experiments and show the way of their
 carrying out. This is the way of a theorist-phenomenologist,
 ``from a phenomenon to a theoretical scheme'' and back.

  At the same time, many considerable results in constructing a
 physical theory have been achieved by another, more speculative way.
 Remember the unification of the force of terrestrial and celestial
 gravity, electricity and magnetism, as well as the recently
 discovered principle of dynamics from symmetry that led to the
 construction to the theory of electro-weak interactions and quantum
 chromodynamics. Adherents of this way, trying to proceed from
 deeper physical representations, initial principles, {\it ab initio},
 are often called \ ``reductionists''. It means that they try to
 reduce the description of the diversity of observed phenomena to
 a small number of simple and general notions and principles.
 In statistical physics \ ``reductionists'' are, as a rule, authors
 of a microscopic approach.

  I would like to cite the definition given by Bogoliubov in 1958 in
 his paper ``Basic principles of the theory of superfluidity and
 superconductivity" \cite{16}:

\begin{quote} {\it
 ``The goal of the macroscopic theory is the derivation of equations
 of the type of classical equations of mathematical physics that
 would reflect the whole set of experimental data related to
 macroscopic objects studied ... .

  The microscopic theory poses a deeper problem of understanding
 the inner mechanism of a phenomenon following the laws of quantum
 mechanics. …  In this case, in particular, one also has to obtain
 relations between dynamic quantities resulting in equations of
 the macroscopic theory''.}
 \end{quote}

 However, one should not take a great accent in opposing these
 two ways of thinking. An important thing is that between the
 equations, for example, classical equations of mechanics or
 Maxwell equations in medium and the laws described by the sequence
 of events such as the laws of planetary motion of the solar system
 or the Meissner law in a superconductor there is an interval, a
 logical gap. These are situations where phenomenology manifests
 itself most strongly. Therefore, the efforts of phenomenologists
 and reductionists supplement each other. The explanation of the
 form and then essence of electro-weak interaction, and the
 phenomena of superfluidity and superconductivity are vivid
 examples of modern quantum theory.

  Bogoliubov and Landau made a pivotal contribution to the creation
 of the theory of macroscopic quantum phenomena, the phenomena of
 superfluidity and superconductivity accompanied by spontaneous
 symmetry breaking at the quantum level.

 \subsubsection{Superfluidity}

 The history of creation of the theory of superfluidity gives a good
 example of interconnection of phenomenological constructions and
 physical ideas. The original explanation of the phenomenon of
 superfluidity given by Landau was based on a general notion that
 at low temperatures superfluid properties of liquid $\,^4He\,$ are
 defined by a linear spectrum of collective excitations (phonons)
 rather than by a quadratic spectrum of excitations of individual
 particles (atoms). It follows from this assumption that in moving
 with velocity not exceeding a certain critical value it is
 impossible to slow down the liquid by transferring energy and
 momentum from the wall to individual atoms because a linear form
 of the phonon spectrum does not allow one to keep simultaneously
 the laws of energy and momentum conservation. The need for
 agreement between the form of the spectrum and the thermodynamic
 properties of liquid helium made Landau introduce excitations,
 in addition to phonons, with a quadratic spectrum beginning with
  a certain energy gap, excitation, which he called rotons.

  Bogoliubov's theory is based on a physical assumption that in
 weakly non-ideal Bose gas there is a condensate akin to ideal
 Bose gas. The existence of the Bose condensate leads to common
 wave function of the whole system, collective effect. Therefore,
 the presence of as weak as is wished interaction transforms
 single--particle excitations into the spectrum of collective
 excitations. To calculate this spectrum, Bogoliubov inferred that
 at low temperatures the Bose condensate contains a macroscopically
 large -- of an order of Avogadro number $N_A$ -- number of
 particles $N_0$, as a result of which matrix elements of the
 creation and annihilation operators of particles in the
 condensate are proportional to square root of "large" number
 $N_0$, and the main contribution to the system dynamics comes
 from the processes of particle transition from the condensate to
 the continuous spectrum and back to the condensate. The
 simplified system of quantum mechanical equations based on
 the afore-said has exact solution and the derived spectrum of
 collective excitations (bogolons) unified phonons and so-called
 Landau rotons. Bogoliubov's courageous intuitive guess of the
 important role of the condensate was experimentally verified
 only thirty years later.

  It is also important that in the Bogoliubov picture there arises
 a natural, though not very transparent, answer to the question
 of the nature of symmetry violated in the phase transition of
 $\,^4He\,$ into a superfluid state. This is phase symmetry of the
 quantum Bose system that (with the help of the Noether theorem)
 is responsible for conservation of the total number of $N$
 particles, i.e., helium atoms in the system considered.
 Collective quasiparticles, bogolons, do not correspond to a
 definite number of $HeII$ atoms representing a superposition
 of an infinite set of particle pairs with zero total momentum.

 \subsubsection{Superconductivity}\label{322}
   Another example of a phase transition in a quantum system
 accompanied by spontaneous symmetry breaking is the phenomenon of
 superconductivity where phase invariance violation occurs, as in
 the case of phase transition to a superfluid state. Though
 superconductivity was discovered in 1911 (much earlier than $^4He\,$
 superfluidity), a theoretical insight into the phenomenon of
 superconductivity was gained much later than explanation of
 superfluidity.

   A breakthrough along this line was a phenomenological theory
 suggested by Ginzburg and Landau (the G-L theory), in which a
 superconducting state was described by an effective complex function
 of a group of electrons playing the role of a two-component order
 parameter. The G-L theory successfully described the behavior of
 a superconductor in the external magnetic field and some other
 important properties. At the same time, the nature of a
 superconducting transition remained unclear.

  The microscopic (quantum) theory of superconductivity was
 developed only in 1957 by Bardeen, Cooper and Schrieffer, and
 Bogoliubov. Three authors have considered a simplified model in
 which an interaction of electrons due to an exchange of phonons is
 substituted for effective attraction of electrons near the Fermi
 surface. The BCS theory includes thermodynamics and electrodynamics
 of a superconductor, calculation of the temperature of a phase
 transition and gives a universal relation between the gap in the
 spectrum at zero temperature and the temperature of a superconducting
 transition. A gap in the spectrum arises due to the formation of
 bound states of electron pairs with the opposite momenta and spins,
 \ ``Cooper pairs'', and is proportional to the exponent
 $e^{-1/\lambda}\,$ where $\lambda$ is the intensity of electron
 attraction.

  Before the appearance of a detailed paper by BCS Bogoliubov
 succeeded in constructing a microscopic theory of superconductivity
 for the complete Fr\"ohlich electron-phonon model. With the new Fermi
 amplitudes he carried out compensation of so-called \ ``dangerous
 diagrams'' corresponding to the production of electron pairs with the
 opposite momenta and spins. The Bogoliubov equations for the energy
 gap and phase transition temperature coincide with the results of
 the BCS theory with the coupling constant $\lambda=g_F^2\,$ directly
  determined by the Fr\"ohlich coupling constant in the
  electron--phonon interaction Hamiltonian.

  Bogoliubov's quasiparticles (sometimes called ``bogolons'') give
 a clear physical picture of the spectrum of quasi-particle excitations
 as a superposition of a particle and a hole that have a gap in the
 spectrum on the Fermi surface\footnote{See our recent review paper
 \cite{60-ufn} on the subject}. Based on the Bogoliubov representation
 of quasiparticles it is easy to calculate thermodynamic and
 electrodynamic properties of a superconductor. The Fermi version of
 the Bogoliubov canonical $(u, v)$ transformation is widely used in
 solving present-day problems in the theory of superconductivity.

   Bogoliubov came to the conclusion of the unity of these two
 macroscopic quantum phenomena: It is superfluidity of Cooper pairs
 that creates a superconducting current. Here is the citation from
 Bogoliubov's review\cite{18} of that time: {\it ``The property of
 superconductivity may be treated as a property of superfluidity of
 a system of electrons in metal.''} The unity of the phenomena of
 superfluidity and superconductivity has recently been confirmed
 in experiments with ultracold fermion gases in traps.

 \section{Scientist and teacher}

 \subsection{Features of Bogoliubov's creativity}

In conclusion, I shall sum up some observations following from the
 analysis of Bogoliubov's scientific creative activity only in
 theoretical physics in the 1950s.

Over that decade NN contributed to about a dozen of scientific
areas:
\begin{center} {\small
\begin{tabular}{|l|c|r|}  \hline
Item & No of papers & period  \\ \hline  \hline
Tomonaga–Schwinger equation with area function  & (4) &1950 – 1952
 \\ \hline
Plasma in magnetic field   &           (8)  & 1951 – 1952  \\ \hline
Functional integral representation  &   (1) & 1954     \\ \hline
Causality condition and scattering matrix  & (3) & 1955 – 1956  \\ \hline
Multiplication of singular functions \& R operation & (5) & 1955 –1957  \\ \hline
Renormalization group    &             (4) & 1955 – 1956  \\ \hline
Physical dispersion relations   &     (3) & 1956 – 1957  \\ \hline
Subtleties of proving DRs  &          (7) & 1956 – 1958  \\ \hline
Superconductivity for Fröhlich model  &  (4) & 1957 – 1958 \\ \hline
Model Hamiltonians and pair correlations & (4)& 1959 – 1960 \\ \hline
Indefinite metric in QFT   &                  (2) & 1958  \\ \hline
Quasi-averages  &                     (2) & 1960 – 1961   \\ \hline
\end{tabular}  }
 \end{center}
  A total of about 50 works and in addition five monographs. It is
 worth mentioning that NN worked on each topic no more than two to
 three years, on the average; in some years he published papers in
 4--5 areas. The especially fruitful period was the mid-1950s.

  Speaking figuratively, in those years Bogoliubov was a fountain
 of fundamentally important scientific discoveries. Benevolence
 towards people, generosity of his nature resulted in that this
 fountain fertilized everybody who had wished to approach it and
 managed to imbibe vivifying water.

 It was just in those years that Nikolai Nikolaevich set up the
 Laboratory of Theoretical Physics as part of the Joint Institute
 for Nuclear Research in Dubna\footnote{The above quotation is
 from my article published with other collected papers \cite{19}
 dedicated to the 50th anniversary of the Bogoliubov Laboratory
 of Theoretical Physics.} and the foundation of his school
 in particle interaction theory was laid.

  For comparison, we may look at such multitalented luminaries as
 Heisenberg and Landau. A cursory glance at their lists of papers
 reveals that each of them returned to the same topic within a
 period of more than a decade.   The motto best suited for N.N.'s
 creative style is \ {\it``Veni, vidi, vici''}. He addressed himself
 to the problem, exhaustively solved it, and switched to another
 problem.

 \subsection{Teacher}

 Unlike Landau, NN never erected a barrier between himself and a
 neophyte in the form of sophisticated entrance examinations. As
 seen from the afore-cited fragments of my formation as a
 scientist, he valued not the initial background but rather the
 ability to enter promptly into the range of new ideas and
 especially the ability to carry on original work. I repeat that
 in the episode with the diploma thesis NN \ ``threw me to learn
 science swimming'' \ straight in deep water. According to legend,
 Rutherford followed the same practice. However, in the case of
 failure, NN did not repudiate the novice but gave him a simpler
 problem to deal with. Partly for this reason the core of Bogoliubov's
 scientific school in quantum field theory formed quite soon, in the
 second half of the 1950s.

 The decisive element of his teaching was scientific generosity:
 the first three papers on the renormalization group in the Doklady
 [6, 8, 12] were published in 1955 under two our names. However,
 having thoroughly analyzed both ultraviolet and infrared asymptotics
 within QED and concluding that argumentation of Landau and
 Pomeranchuk [3] on the zero charge lacked proving power, NN's
 interest in the renormalization group slightly cooled down and,
 assigning me a task in the meson–nucleon theory, he switched to
 other problems (see the previous section). He decidedly refused to
 be a co-author of the next publication \cite{20}\footnote{Already
 the next year I applied this \ ``minimum co-authorship'' \ rule to
 diploma work of my first graduation students Ilya Ginzburg and Lev
 Soloviev.}.

  Another teaching method was involvement of a young colleague into
 a large activity such as joint work on a book. Finally, the third
 method of cultivating independence consisted in accelerated
 training of the young co-author in the art of reporting on joint
 research. For example, apart from seminars, I had to make a
 review report \cite{21} at the 3rd All-Union Mathematical Congress
 in Moscow in 1956 and a rapporteur's report \cite{22} at the
 Rochester HEP Conference in Kiev in 1959. In the latter case, the
 replacement of the speaker came as a surprise from N.N. just
 the day before.

   The above example of minimum co-authorship illustrates a
 constituent of scientific scrupulousness of Nikolai Nikolaevich.
 As a second constituent, I would mention his high (first seemed
 unreasonable) demands for thorough citing of predecessors in one
 or another scientific topic. Finally, the third constituent is
 responsibility for literally each line in the text of paper.

  My long experience of co-authorship with N.N. has made me inclined
 (which is sometimes noticeably burdensome for me and my co-authors)
 to clear comprehension and maximum clarity of the formulations in
 reasoning and results as well as to the clear indication of the
 reasons why this or that paper is mentioned\footnote{The
 consequences are dislike to \ ``common graves'' \ in citing and
 cases of leaving the team of authors because of disagreement on an
 important element of the description of the joint research.}.

  Finally, a few words about human scrupulousness. I do not remember
 a case where I had to experience any pressure from Bogoliubov though
 he was not only the scientific leader but also the higher--ranking
 administrative person. N.N. usually just only offered a scientific
 idea or some practical solution to the colleague or subordinate. He
 did it gently and never insisted if no positive response was given.
 ``From each according to his abilities''. This happened to me when he
 looked for an assistant in organizational efforts during the
 establishment of the Theoretical Laboratory at JINR in 1956 and when
 I came to LTP again in the early 1970s. This happened many times
 with me and his other students in scientific topics. NN usually
 foresaw the way in which topical scientific idea would develop and
 advised his co-workers to deal with one or another issue.
 Remembering these cases, we regret that being carried away at that
 moment by something else (less important, as time showed), we paid
 no due heed to his recommendations.

  Bogoliubov was not indifferent to personal qualities of young
 people whom he favored. NN valued the congenial human environment,
 the moral climate among his co-workers. My memory keeps two cases
 of ostracism. One of them concerned the then young scientist Y,
 very gifted but already too free and easy with the colleagues
 (working on similar problems). A glance at the list of NN's works
 shows that he often got several people involved in solving a problem.
 Friendly relations between them were a norm. In the case with Y,
 however, after several conflicts the people had to turn to the chief.
 And he debarred Y from the work. The second episode involved the
 older colleague Z with a difficult biography warped by repressions
 of the 1930s. Once during a regular tightening of ideological nuts
 in the 1970s Z became a witness of \ ``seditious'' \ political
 statements in a not very narrow circle of the Fizfak staff members.
 Fearing than any of other witnesses could report it to the State
 Security, he did it himself. And this became officially known. The
 chief was quick to react. Remarkably, N.N. understood that action
 from the pure human point of view. He understood, felt sympathy
 inwardly, and explained the reasons through an old Indian parable.
 But he did not ever want to deal with this man with whom he
 fruitfully collaborated for about 20 years. Most important moral
 lessons of Nikolai Nikolaevich were learnt not from his reprimands
 but rather his behavior and the way of actions. So his ability to
 combine scientific work with civic duty, including scientific and
 administrative positions, served as an example for some of his
 outstanding disciples.

 This became apparent in a very complicated post-Soviet period.
 Unlike many prominent Soviet scientists, representatives of the
 Bogoliubov School served and will continue serving their Motherland.
 Thanks to them the spirit of Nikolai Nikolaevich is still among us.

 The paper was partially supported by grant NSh-1027.2008.2.
 

\small
 \begin{thebibliography}{99}
 \bibitem{1} N.N. Bogoliubov and D.V. Shirkov, {\it Uspekhi Fiz. Nauk}
 \ {\bf 55}, 149-214 (1955).
\bibitem{2} N.N. Bogoliubov and D.V. Shirkov, {\it Uspekhi Fiz. Nauk}
 \ {\bf 57}, 3-92 (1955).
\bibitem{3} L.D. Landau and I.Ya. Pomeranchuk,
          {\it Doklady Akad. Nauk SSSR} \  {\bf 102 }, 489 (1955).
\bibitem{4} M .Gell-Mann and F.Low, Phys. Rev. 95, 1300 (1954).
  \bibitem{5} E.C.G. Stueckelberg and A. Petermann, ``La normalisation
 des constants dans la theorie des quanta'', {\it Helv. Phys. Acta}
  \ {\bf 26} 499–520 (1953) --in French.
 \bibitem{6} N.N. Bogoliubov and D.V. Shirkov, ``On the
 Renormalization Group in Quantum Electrodynamics'', {\it Doklady Akad.
 Nauk SSSR} {\bf 103},  203 (1955) --in Russian.
 \bibitem{7} N.N. Bogoliubov and D.V. Shirkov, "Charge Renormalization
 Group in Quantum Field Theory", Nuovo Cim. 3, 845-863 (1956).
 \bibitem{8} N.N. Bogoliubov and D.V. Shirkov, ``Use of the
  Renormalization Group For Improving Perturbation Theory Formulas'',
 {\it Doklady Akad.Nauk SSSR} {\bf 103}, 391--394 (1955) --in Russian.
 \bibitem{9} N.N. Bogoliubov, {\it J. Phys. USSR} \ {\bf 11}, 23-32
 (1947) [Submitted 12 X 1946]\footnote{Numbers in brackets are the
 date of submitting; numbers in parentheses are the date of publication.}.
 \bibitem{10} L.D. Landau, {\it J. Phys. USSR} {\bf 11}, 91-92 (1947)
  [15 XI 1946].
 \bibitem{11}L.D. Landau, {\it Doklady Akad. Nauk SSSR} \ {\bf 61},
  253 (1948) [15 VI 1948].
\bibitem{12}L. D. Landau, {\it Phys. Rev.} \ {\bf 75}, 884 (1949).
\bibitem{13} N.N. Bogoliubov and D.V. Shirkov, ``Model of the Lie type
 in QED'', {\it Doklady Akad. Nauk SSSR} \ {\bf 105}, 685-688 (1955).
 \bibitem{14}L.D. Landau, ``On the Quantum Theory of Fields'' in
 {\sf ``Niels Bohr and the Development of Physics''}, eds. W.Pauli et
 al, Pergamon, London, 1955.
  \bibitem{L-C} L.N. Cooper {\it Phys. Rev.} {\bf 104}, 1189 (1956).
 \bibitem{15} N.N. Bogoliubov, {\it Zh.Eksp.Teor. Fiz.} \ {\bf 34}, 58
  (1958); \ {\it Nuovo Cim.} {\bf 7}, 794 (1958).
 \bibitem{60-ufn} D.V. Shirkov, ``60 years of Broken Symmetries in
 Quantum Physics: From the Bogoliubov Theory of Superfluidity to the
 Standard Model'' Phys.Usp.{\bf 52}:549-557,2009 ; arXiv:0903.3194
 [physics.hist-ph].
 \bibitem{16} N.N. Bogoliubov, {\it Vestnik Akad. Nauk SSSR}, No. 4, 25
  (April 1958) -- in Russian; see pp. 289--296 in [17].
 \bibitem{17} N. N. Bogoliubov, {\sf Collected Works in 12 volumes},
 Vol. VIII, Moscow, Nauka, 2008 -- in Russian.

  \bibitem{18} N. N. Bogoliubov, {\it Vestnik Akad. Nauk SSSR}, No. 8,
   36 (August 1958)-- in Russian; see also pp. 289--296 in [17].
  \bibitem{19} {\sf Collected Papers ``LTPh 50''}, Dubna, JINR Publ.
   Dep., 2006.
  \bibitem{20} D.V. Shirkov, Doklady Akad. Nauk SSSR  105, 972–975(1955).
 \bibitem{21} N.N. Bogoliubov and D.V. Shirkov, "Some Issues of QFT",
 {\sf Proc. 3rd All-Union Mathematical Congress}, Vol. 2, Moscow,
 Acad. Sci. USSR Publ., 1956, pp. 84--85 -- in Russian.
 \bibitem{22} D.V. Shirkov, ``Theoretical Studies of Dispersion
 Relations'' in {\sf Proceedings of 9th Int. Conf. High-Energy Phys.},
  Vol. 2, Kiev, 1956, pp. 3--22.
\end{thebibliography}
\end{document}